\documentclass{PoS}

\usepackage{graphics}
\usepackage{graphicx}
\usepackage{epsfig}
\usepackage{amssymb}

\newcommand{\be}{\begin{equation}}
\newcommand{\ee}{\end{equation}}
\newcommand{\ba}{\begin{eqnarray}}
\newcommand{\ea}{\end{eqnarray}}

\def\lsi{\raise0.3ex\hbox{$<$\kern-0.75em\raise-1.1ex\hbox{$\sim$}}}
\def\gsi{\raise0.3ex\hbox{$>$\kern-0.75em\raise-1.1ex\hbox{$\sim$}}}
\newcommand{\lsim}{\mathop{\lsi}}
\newcommand{\gsim}{\mathop{\gsi}}
\newcommand{\eq}{Eq.~}

\newcommand{\fig}{Fig.~}

\title{Exploring the QCD phase diagram}

\ShortTitle{Exploring the QCD phase diagram}

\author{\speaker{Owe Philipsen}\\
        Universität Münster\\
        E-mail: \email{ophil@uni-muenster.de}}

\abstract{
Lattice simulations employing reweighting and Taylor expansion 
techniques have predicted a $(\mu,T)$-phase diagram
according to general expectations, with an analytic quark-hadron crossover at $\mu=0$
turning into a first order transition at some critical chemical potential $\mu_E$.
By contrast, recent simulations using imgainary $\mu$ followed by
analytic continuation obtained a critical structure in the $\{m_{u,d},m_s,T,\mu\}$ parameter space
favouring the absence of a critical point and first order line. 
I review the evidence for the latter scenario, arguing that the various raw data are not inconsistent
with each other. Rather, the discrepancy appears when attempting
to extract continuum results from the coarse ($N_t=4$) lattices simulated so far, and can be explained
by cut-off effects. New (as yet unpublished) data are presented, which for $N_f=3$ and on
$N_t=4$ confirm the scenario without a critical point. Moreover, simulations on finer $N_t=6$ lattices
show that even if there is a critical point, continuum extrapolation moves it to significantly 
larger values of $\mu_E$ than anticipated on coarse lattices.
} 

\FullConference{Critical Point and Onset of Deconfinement   -  4th
International Workshop\\
		 July 9 - 13, 2007\\
		 Darmstadt, Germany}

\begin{document}

\section{Introduction}

\begin{figure}[t]
\vspace*{-0.5cm}
\centerline{
\scalebox{0.58}{\includegraphics{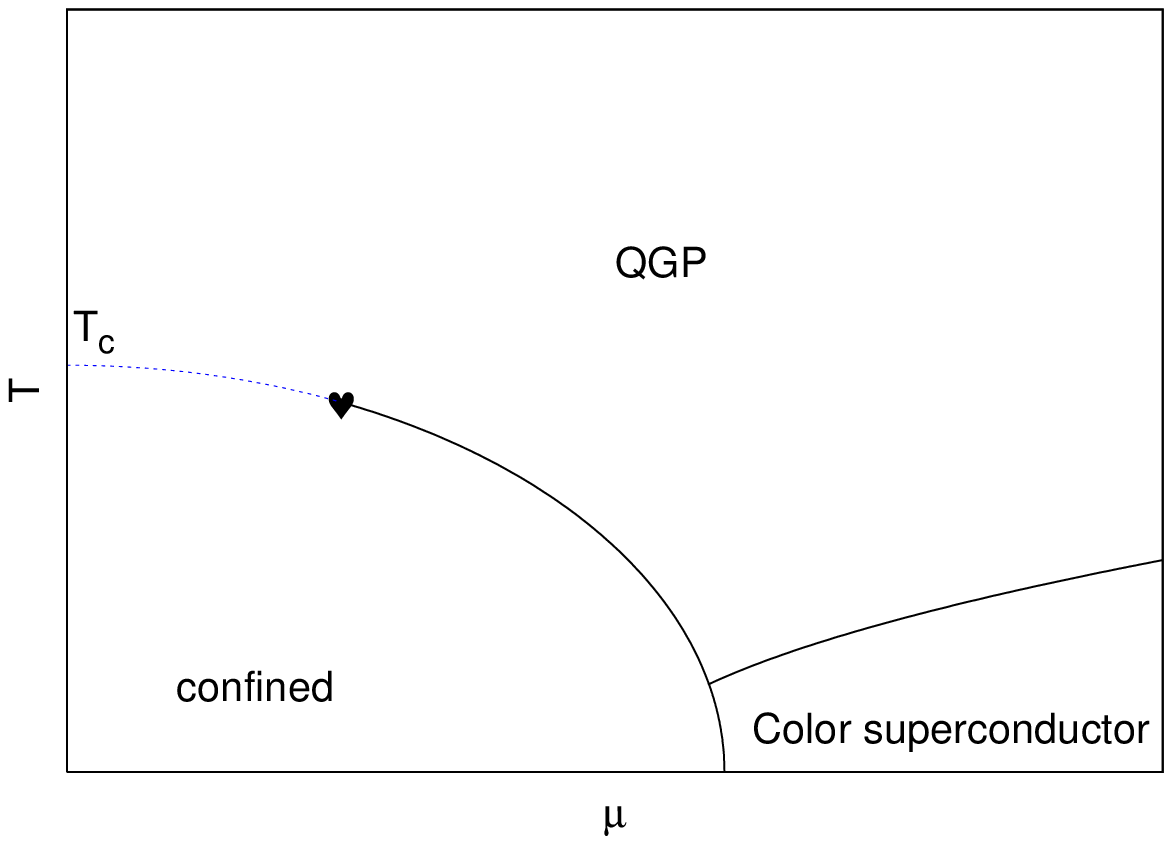}}
\scalebox{0.58}{\includegraphics{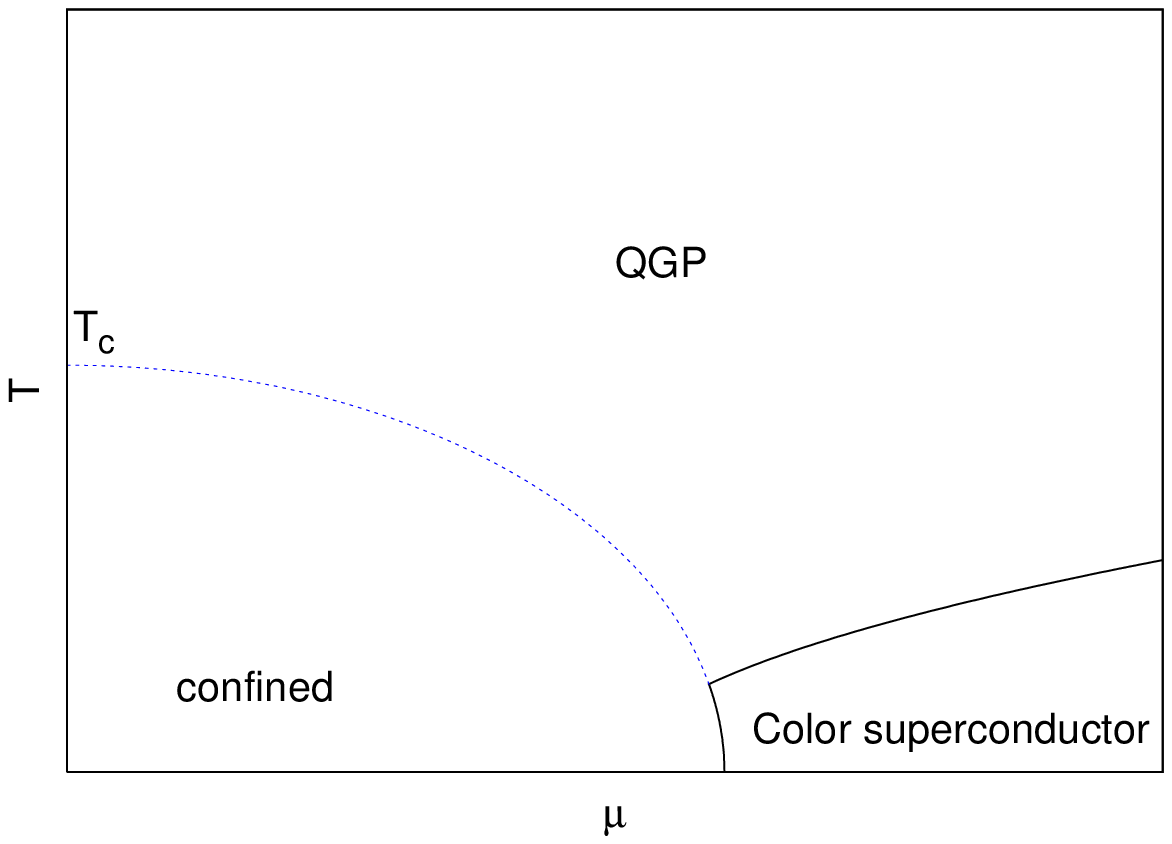}}
}
\caption[]{Left: Expected phase diagram for physical QCD. Right: The prediction from a recent lattice
investigation \cite{fp3}. The entire phase boundary between plasma and confined phase corresponds to a crossover.}
\label{pd}
\end{figure}
One of the most fascinating and challenging tasks in the study of QCD under 
extreme conditions remains to elucidate its phase diagram as a function of 
temperature and baryon density. QCD is a strongly coupled theory, and it has recently 
become clear that certain strong coupling features
persist also in the quark gluon plasma regime up to the temperatures that are
currently accessible experimentally. Hence, perturbation theory fails us for this problem, and 
simulations of lattice QCD are the only known tool by which we may hope to ultimately come to reliable
theoretical predictions for QCD. Unfortunately, until 2001 the so-called ``sign-problem'' of 
lattice QCD has prohibited simulations at finite baryon density.   

Despite this lack of reliable calculations, it is generally believed that the phase diagram qualitatively looks as in \fig\ref{pd} (left). 
Lattice simulations have been operating along the temperature axis at $\mu=0$ for a long time, 
and results have been consistent with an analytic crossover in this regime. However, it was only
very recently that these computations could be performed with realistic quark masses
and on reasonably fine lattices \cite{fknat}, so that we now may have confidence in this prediction.
For $T=0$ and finite density, on the other hand, a number of models and perturbative arguments 
\cite{wilc,tric} predict a first order phase transition to a superconducting regime at some $\mu_c\gsim 1$ GeV.
In the simplest scenario the phase boundary encountered in this finite density transition and that of the finite temperature transition are continuously connected. 
In this case the first order transition line must terminate in a second order end-point, thus leading to the familiar phase diagram, \fig\ref{pd} (left). 
For more detailed arguments and references, see e.g.~\cite{wilc,misha}. 

Significant progress in lattice QCD was made since 2001, with several different
methods now available that circumvent the sign problem,
rather than solving it: i) Multi-parameter reweighting \cite{fk1}, 
ii) Taylor expansion around $\mu=0$ \cite{swa} and iii) simulations at imaginary 
chemical potential, either followed by analytic continuation \cite{fp1} or Fourier transformed to the canonical ensemble \cite{slavo}. 
It is important to realize that all of these approaches introduce some degree of 
approximation. However, their systematic errors are rather different, thus allowing for powerful crosschecks.  
All methods are found to be reliable as long as $\mu/T\lsim 1$, or $\mu_B\lsim 550$ MeV, which includes the region of interest for heavy ion collisions. 
Reviews specialized on technical aspects can be found in \cite{oprev,csrev}. 

In particular, a calculation using reweighting methods for $N_f=2+1$ with 
physical quark masses on a coarse lattice ($N_t=4$, $a\sim 0.3$ fm) appeared to confirm the 
expected phase diagram and gave a prediction for the critical end point at $\mu_B^E\sim 360(40)$ MeV \cite{fk2}. 
Similarly, an investigation of the convergence radius
for the Taylor series of the pressure in the $N_f=2$ theory with bare quark mass $m/T=0.1$ yielded a signal for the endpoint at $\mu^E_B/T_E=1.1\pm 0.2$, 
$T_E/T_c(\mu=0)=0.95$ \cite{ggpd}.
By contrast, a recent study of $N_f=2+1,3$ using imaginary chemical potentials reached a surprising conclusion 
contradicting the standard scenario: for physical values of the quark masses there would 
be no critical point or first order line at all, the crossover region extends all the way to a possible
line delimiting the superconducting phase, as in \fig\ref{pd} (right) \cite{fp3}. 
In the following I will go through the evidence for this
unorthodox scenario and argue that the raw data of the apparently contradicting investigations are
actually consistent with each other. The apparent discrepancy arises when
attempting to extract continuum physics and can be explained by standard cut-off effects.

\section{The nature of the QCD phase transition: qualitative picture}

\begin{figure}[t]
\vspace*{-5.5cm}
\includegraphics[width=0.5\textwidth]{phase_diagram_trunc.ps}
\includegraphics[width=0.5\textwidth]{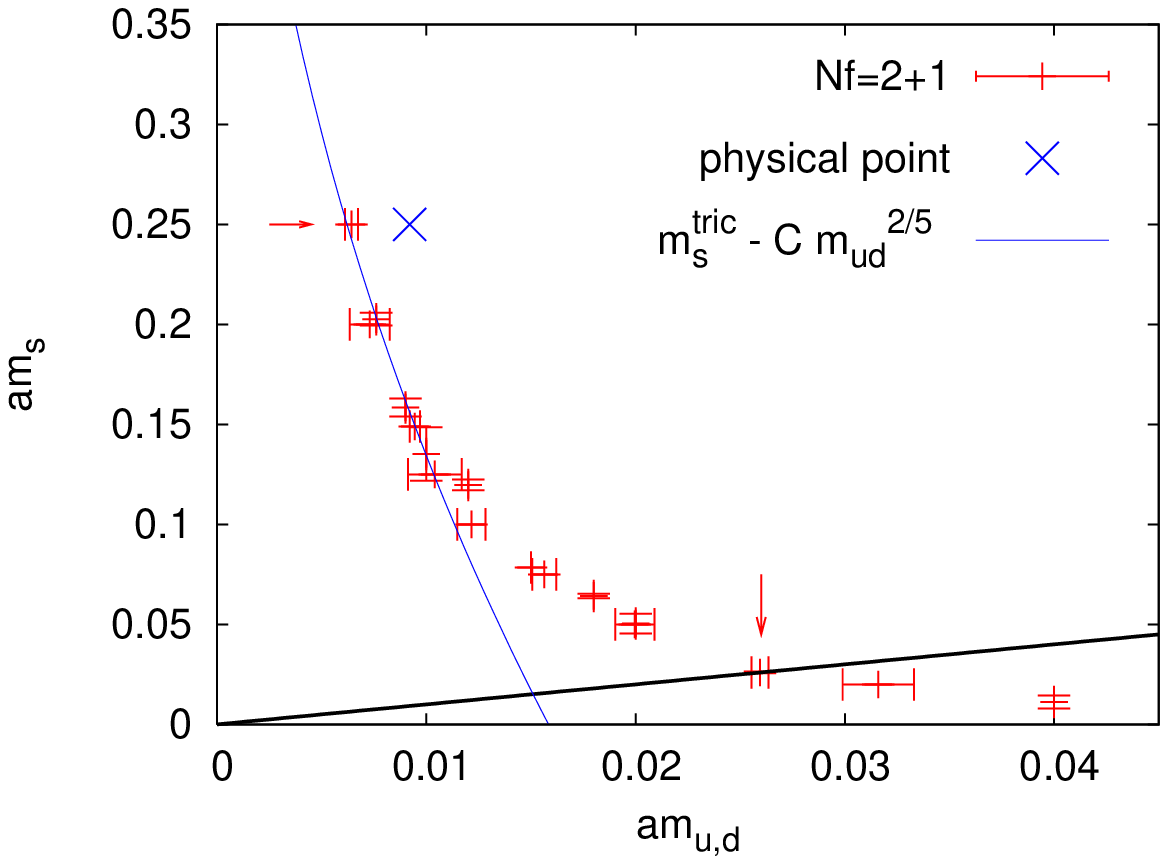}
\caption[]{Left: Schematic phase transition behaviour of $N_f=2+1$ flavor QCD for
different choices of quark masses $(m_{u,d},m_s)$, at $\mu=0$ Right: Numerical results for the chiral critical line. Also shown are the physical point according to
\cite{fk2}, and a fit corresponding to a possible tricritical point $m_s^{tric}\sim 2.8 T$.}
\label{m1m2c}
\end{figure}

Rather than fixing QCD to physical quark masses and just varying $T,\mu$, it is instructive to consider
varying quark masses and the full parameter space $\{m_{u,d},m_s,T,\mu\}$. The qualitative picture for
the order of the phase transition at zero density is shown in
\fig\ref{m1m2c} (left).
In the limits of zero and infinite quark masses
(lower left and upper right corners),
order parameters corresponding to the breaking of a global symmetry can be defined,
and one finds numerically at small and large quark masses
that a first-order transition takes place at a finite
temperature $T_c$. On the other hand, one observes an analytic crossover at
intermediate quark masses. Hence, each corner must be surrounded by a region of
first-order transition, bounded by a second-order line. 

\begin{figure}[t!]
\vspace*{-1cm}
\centerline{
\scalebox{0.65}{\includegraphics{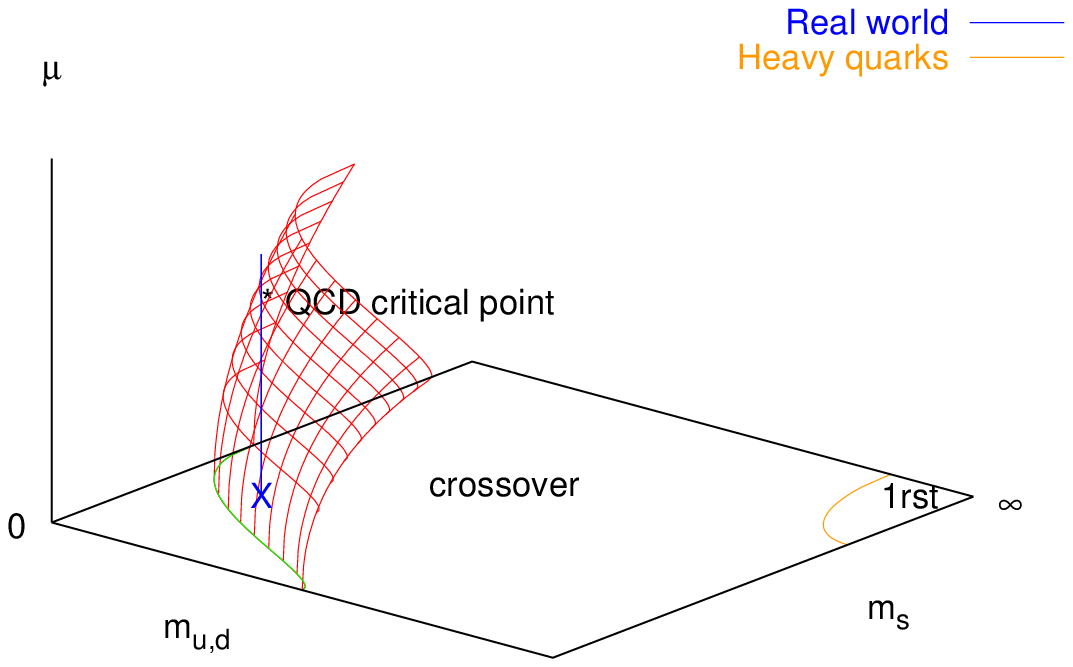}}
\scalebox{0.65}{\includegraphics{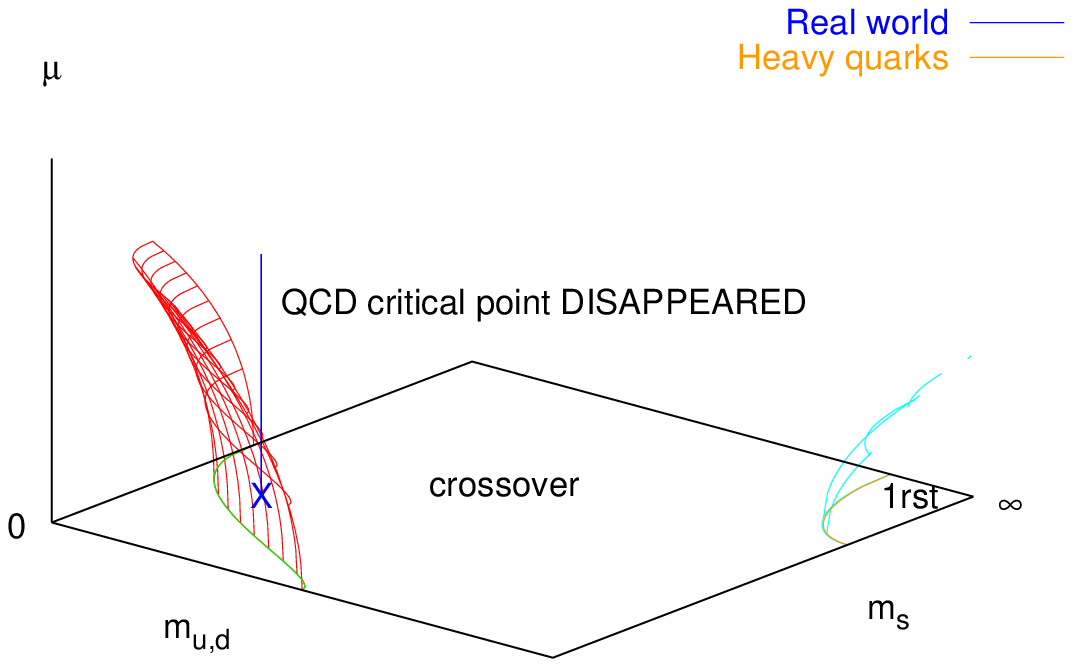}}
}
\vspace*{-0.5cm}
\caption{ 
The chiral critical surface in the case of positive (left) and negative (right) curvature.
If the physical point is in the crossover region for $\mu=0$, a finite $\mu$
phase transition will only arise in the scenario (left) with positive curvature,
where the first-order region expands with $\mu$.
Note that for heavy quarks, the first-order region shrinks with $\mu$ \cite{Potts}.
}
\label{2schem}
\end{figure}

Now consider the effect of a baryonic chemical potential, $\mu_B=3\mu$.
As a function of quark chemical potential $\mu$, represented vertically in \fig\ref{2schem}, 
the critical line determined at $\mu=0$ now spans a surface. The standard expectation for the QCD
phase diagram is depicted in 
\fig\ref{2schem} (left). The first order region expands as $\mu$ is turned on, so that the
physical point, initially in the crossover region, eventually belongs to the
critical surface. At that chemical potential $\mu_E$, the transition is second order:
that is the QCD critical point. Increasing $\mu$ further makes the transition
first order. Drawn in the $(T,\mu)$-plane, this corresponds to the
expected diagram \fig\ref{pd} (left).
A completely different scenario arises if instead the first-order
region shrinks as $\mu$ is turned on, \fig\ref{2schem} (right). Now the physical
point remains in the crossover region for any $\mu$. Hence, existence and location of the critical 
point depend on two parameters of this phase diagram: the distance of the physical point 
from the critical line at $\mu=0$, and the sign and strength of the critical surface's bending.

\subsection{How to ``measure'' the properties of a phase transition}

\begin{figure}[t]
\includegraphics[width=0.3\textwidth]{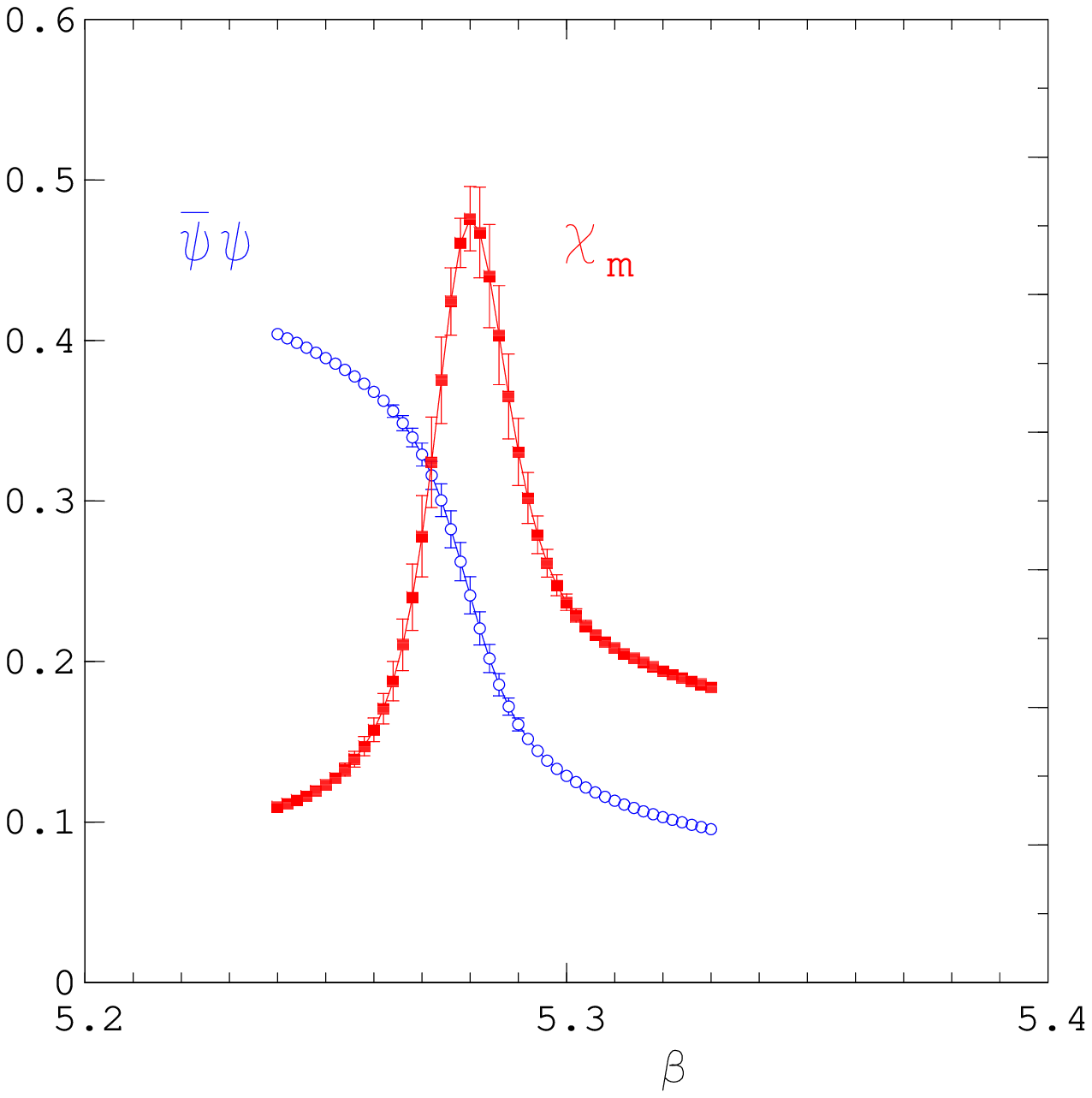}\hspace*{2.5cm}
\includegraphics[width=0.55\textwidth]{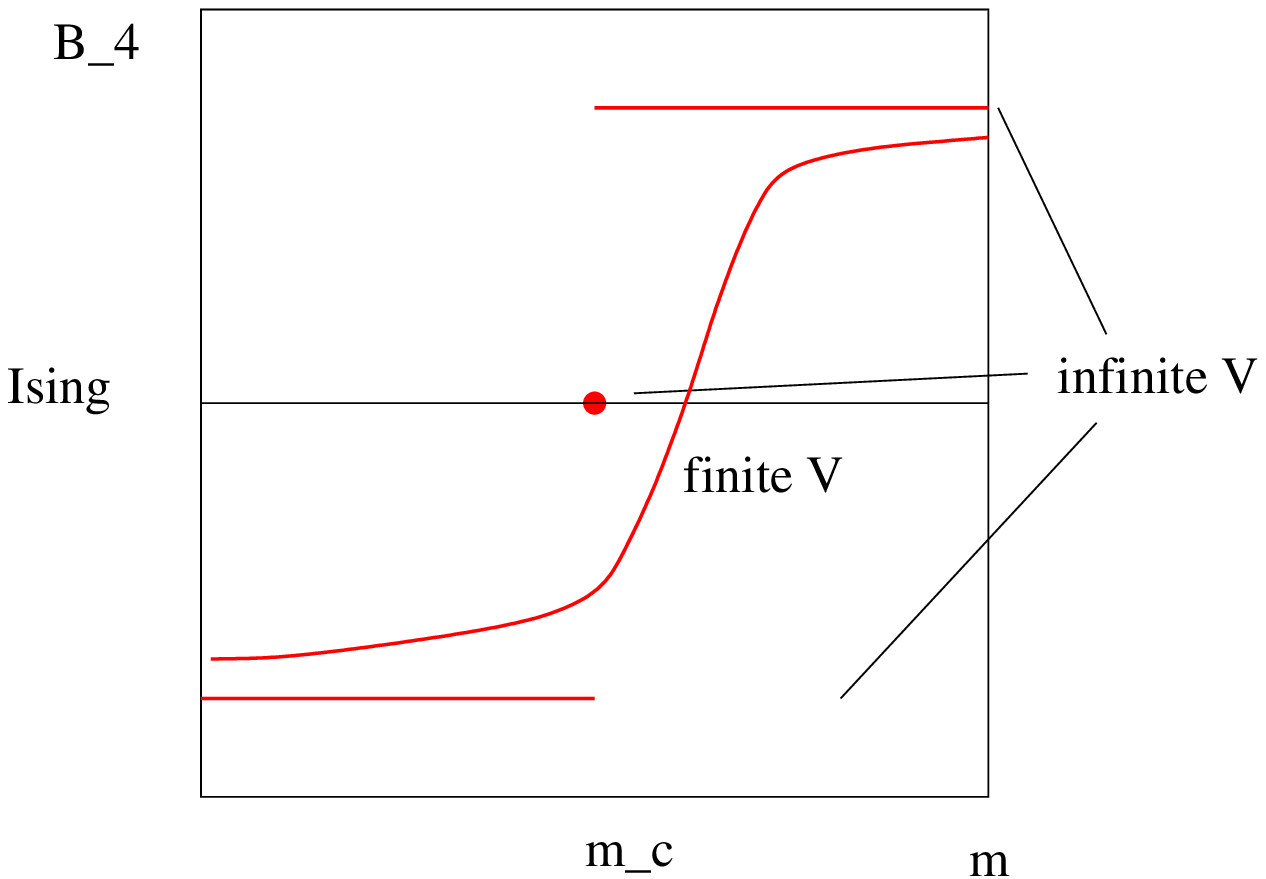}
\caption[]{Left: The chiral condensate and its susceptibility when passing through the (pseudo-)critical coupling of the phase boundary. Right: Finite size behaviour of the Binder cumulant around a 3d Ising
critical point. }
\label{order}
\end{figure}

A lattice calculation of the properties of a phase transition consists of two steps. First, one needs to find the phase boundary, i.e.~the critical coupling $\beta_c$ separating the confined/chirally broken regime
from the deconfined/chirally symmetric regime. To this end one computes the expectation value of the ``order parameters'' $O$ for those symmetries, like the chiral condensate $\langle \bar{\psi}\psi\rangle$ or the expectation value of the Polyakov loop, as well as its associated fluctuations, or generalized susceptibilities, $\chi=V(\langle O^2\rangle -\langle O \rangle ^2)$.
For fixed quark masses, the transition is then announced by the rapid change of the order parameter
as a function of lattice gauge coupling $\beta$, and the peak of the susceptibility 
serves to locate it more precisely, cf.~\fig\ref{order} (left). 
Using the two-loop beta-function (or some numerical
fit to a non-perturbative beta-function), $\beta_c$ 
can be converted to a critical temperature $T_c$. 
This first step of locating the phase boundary is
comparatively easy.

However, on a finite volume, this critical coupling, viz.~temperature, is only pseudo-critical, no true
non-analytic phase transition can occur on finite volumes. The order of the transition thus has to be determined by a finite size scaling analysis. With larger and larger volumes, the peak in the
susceptibility will begin to diverge if there is a true phase transition, and the rate of its divergence, associated with some critical exponent, determines whether it is a first or second order phase transition.
In the case of a crossover the peak stays finite and analytic.
Another, for our purposes more practical, observable is the Binder cumulant, to be evaluated at 
$\beta=\beta_c$, 
\be
B_4(m,\mu)=\frac{\langle(\delta\bar{\psi}\psi)^4\rangle}
{\langle(\delta\bar{\psi}\psi)^2\rangle^2},\qquad \delta \bar{\psi}\psi=\bar{\psi}\psi-\langle \bar{\psi}\psi\rangle.
\ee
In the infinite volume limit, $B_4\rightarrow 1$ or $3$ for a first order transition or crossover, respectively, whereas it assumes a particular value characteristic of the universality class at a critical point. 
For 3d Ising universality, one has $B_4\rightarrow 1.604$.
Hence $B_4$ is a non-analytic step function, which gets smoothed out to an 
analytic curve on finite volumes,
its slope increasing with volume to gradually approach the step function, \fig\ref{order} (right).
Once the universality class of a critical point is established by finite size scaling, 
$B_4$ allows to investigate the change of the order of a transition 
already on a single (large enough) volume.
In \cite{fp3}, we have used $8^3, 12^3$ and $16^3 \times 4$ lattices with standard staggered fermions
to estimate the critical couplings for which $B_4 = 1.604$, and check their consistency on different volumes. 
Here, I also report on new
$\mu=0$ simulations on $12^3,18^3\times 6$.

\section{The chiral critical line $m_s^c(m_{u,d})$}

For $N_f=3$, and at zero density, the critical quark mass marking the boundary between the crossover and the first order region is large enough for simulations to be carried out, and it was possible to determine it with some accuracy \cite{kls,clm,fp2,fp3}. Moreover,  
finite size analyses established it to belong to  
the Z(2) universality class of the 3d Ising model \cite{kls}. 
The critical mass can be read off from the red squares in \fig\ref{mc} (left), 
where data for the Binder cumulant are
shown as a function of varying quark mass, $am^c=0.0263(3)$ or $m^c/T_c=0.105(1)$ \cite{fp3}.

Starting from this point, the boundary line
has been mapped for the non-degenerate $N_f=2+1$ theory, 
again with $N_t=4$ \cite{fp3},
where we have assigned a chemical potential to the two degenerate light quarks only. 
The methodology is the same as in the three flavour case: 
fix a strange quark mass $am_s$ and
scan the Binder cumulant in the light quark mass $am_{u,d}$ for the 
corresponding critical point.
Repeating for several strange quark masses produces the critical 
line $am_s^c(am_{u,d})$ shown in \fig\ref{mc} (right). 

In order to set a physical scale for the boundary line, spectrum 
calculations at $T\sim 0$ have been performed \cite{fp3} 
at the parameters indicated by the arrows
in \fig \ref{mc} (right). They show that 
$m_s$ at the upper arrow is approximately tuned to its physical value 
($\frac{m_K}{m_\rho} \sim \frac{m_K}{m_\rho}|_{\rm phys}$=0.65),
while the pion is lighter than in physical QCD 
($\frac{m_\pi}{m_\rho}=0.148(2) < \frac{m_\pi}{m_\rho}|_{\rm phys}=0.18$).
This confirms that the physical point lies to the right of the critical line,
i.e. in the crossover region.  

\section{Strong cut-off effects on the critical line}

\begin{figure}[t]
\includegraphics*[width=0.5\textwidth]{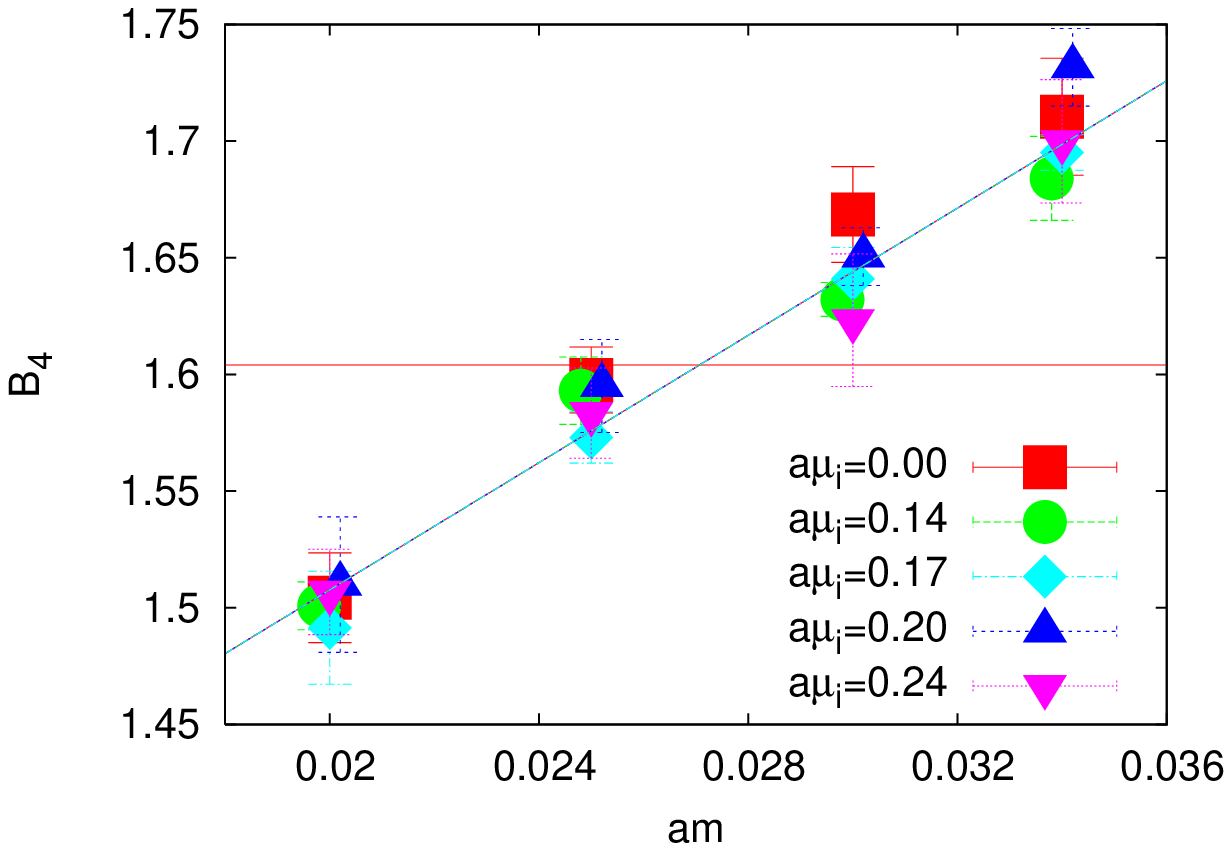}\hspace*{1cm}
\includegraphics*[width=0.5\textwidth]{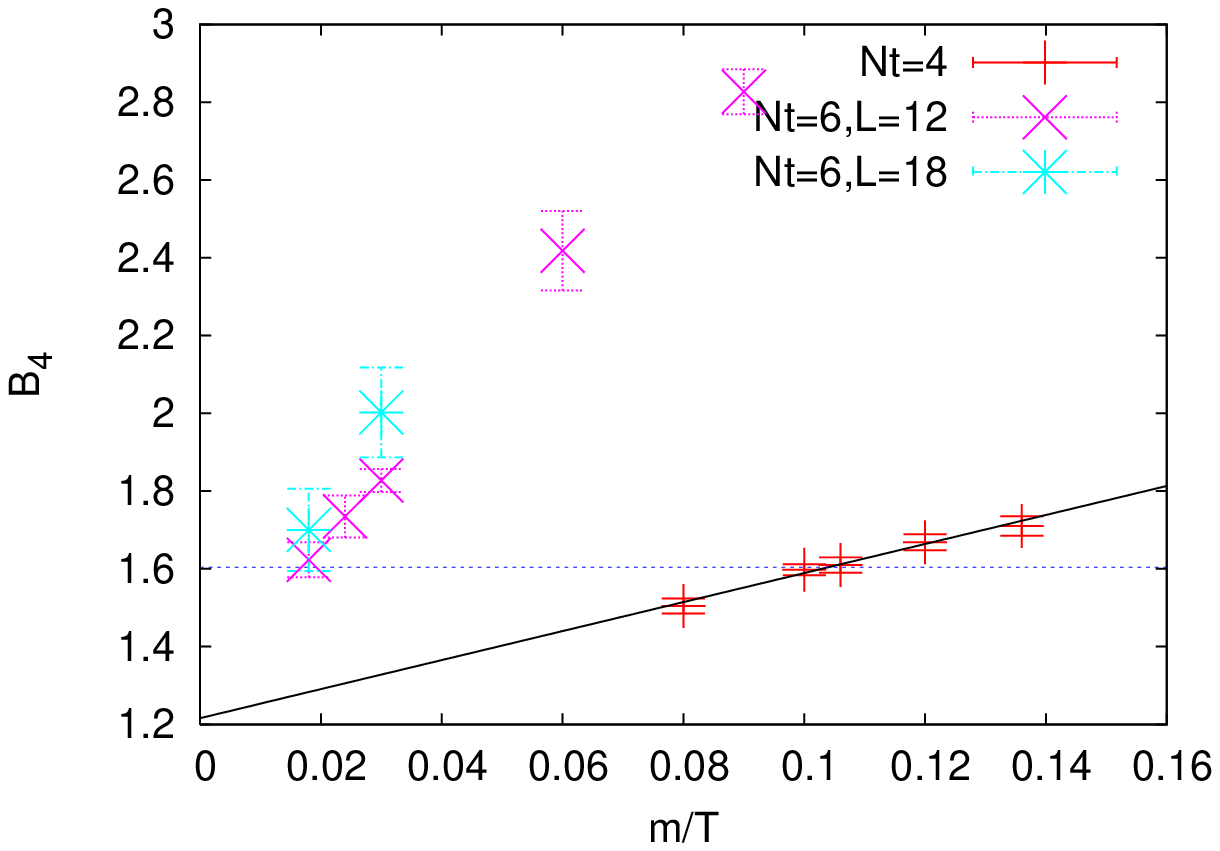}
\caption[]{Left: $B_4$ as a function of quark mass and $\mu_i$ for $N_f=3,N_t=4$.
Right: Comparison of the Binder cumulant for $\mu=0$ on $N_t=4,6$ lattices.}
\label{mc}
\end{figure}

While the universality class of a critical point is determined 
by long range fluctuations and thus should be insensitive to cut-off effects 
on a coarse lattice, the critical quark mass is a
quantity requiring renormalization and expected to be highly sensitive. 
Here we present as yet unpublished data of a simulation on
the finer $N_t=6$ lattice for $N_f=3$. \fig\ref{mc} (right) shows a comparison between the
$N_t=4$ and $N_t=6$ data for $B_4(m)$, as a function of $m/T$. If the lattice spacings were in the
scaling region, the critical values extracted from the intersection with the Ising value should be
close to each other. In contrast, the critical bare quark mass on the $N_t=6$ lattice is 
smaller by almost a factor of five, $m^c/T\approx 0.02$. 
We have also computed the pion mass corresponding to this $m^c$ on $N_t=4,6$, 
finding $m_\pi^c\approx 1.68, 0.96$, respectively. 
Similar observations have been made when comparing
unimproved and improved staggered fermion results on $N_t=4$ lattices \cite{kls}.
Hence,  the boundary line in the phase diagram appears to be still far from the continuum result.
Note that this implies the gap between the physical point and the critical line to widen significantly
towards the continuum, thus moving $\mu_E$ to larger values for fixed curvature of the critical surface. 

\section{Finite density phase diagram using imaginary $\mu$}

For imaginary $\mu=i\mu_i$ the fermion determinant is real positive and 
simulations can be carried out in the same way as at $\mu=0$. The problem then is to get back to 
the physical situation or real $\mu$. This is easily possible if an observable, measured as a function of imaginary chemical potential, can be fitted by a truncated Taylor expansion in $\mu_i/T$, 
\be
\langle O\rangle = \sum_n^N c_n \left(\frac{\mu_i}{\pi T}\right)^{2n}.
\ee
Note that, because of an exact reflection symmetry of QCD under $\mu\rightarrow -\mu$, only even terms 
appear. Clearly, $\mu/(\pi T)$ has to be small enough for this series to converge well, but there
is a limited radius of convergence: when going in the imaginary direction, a Z(3) transition to the neighbouring, and unphysical, center-sector of the gauge group is crossed at $\mu_i/T=\pi/3$, thus 
limiting the useful range for this approach to $|\mu|/T\lsim 1$.  

%
%

\subsection{The critical temperature at finite density}

The peak position in the susceptibilities calculated for several imaginary
$\mu$ can be 
fitted by a leading $O(\mu^2)$ and 
next-to-leading $O(\mu^4)$ Taylor series,  
the data are well described by the leading term, 
$\beta_c(am, a\mu) = 5.1369(3) + 1.94(3) (a m - a m^c_0) + 0.781(7) (a \mu)^2+\ldots$  
Converting to continuum units by means of the two-loop beta-function yields the result
\be
\frac{T_c(\mu,m)}{T_c(\mu=0,m_c(0))}=1 +2.111(17)\frac{m-m_c(0)}{\pi T_c}
-0.667(6)\left(\frac{\mu}{\pi T_c(0,m)}\right)^2+\ldots
\ee
Note that one obtains coefficients of order one when expanding in the ``natural'' parameter 
$(\mu/\pi T)$, as one might expect on the grounds that the Matsubara modes of finite temperature
field theory, coming in multiples of $\pi T$, are setting the scale of the problem.
The coefficients depend on $N_f$, a comparison of the leading $\mu^2$-coefficient for
various cases can be found in \cite{csrev}. 
The curvature gets stronger with increasing $N_f$, which is 
consistent with expectations based on the large $N_c$ expansion.
Subleading coefficients are also beginning to emerge as we discuss in section \ref{check}. 
Note that the continuum conversions using the two-loop beta-function 
are certainly not reliable for these coarse lattices, while fits to 
non-perturbative beta-functions tend to increase the curvature.
Finally, detailed comparisons of $\beta_c(\mu)$ calculated via reweighting techniques
or analytic continuation have been made for $N_f=2,4$ using equal lattice actions, and complete
quantitative agreement is found \cite{slavo}.

\subsection{The critical point as a function of quark masses}

Let us now move to the second step in the program and discuss the order of the phase transition
as a function of quark masses and chemical potential.
The strategy followed in \cite{fp3} is to compute the change of the critical line as a function of imaginary chemical potential and fit it by a truncated Taylor series, just as was done for the critical coupling.
Again the Binder cumulant was used to determine the order of the transition.
The data for $N_f=3$ and various $\mu_i$ are shown in \fig\ref{mc} (left). 
The chemical potential is found to have almost no influence on $B_4$ as a function
of quark mass. The change of $m^c$ with $\mu^2$, i.e.~the leading coefficient in a Taylor expansion, 
is determined from
$d(am^c)/d(a\mu)^2=-(\partial B_4/\partial (a\mu)^2)/(\partial B_4/\partial (am))$ and we obtain
\be
a m^c(a \mu) = 0.0270(5) - 0.0024(160) (a \mu)^2.
\label{c_prime1}
\ee

Care must be taken for the conversion to physical units. The crucial
point is that $T=1/(aN_t)$, and for fixed $N_t$ the lattice spacing is adjusted with 
changing $\mu$ to tune the temperature so as to stay on the critical line, i.e.~$a(T(\mu))$ changes.
Expressing the change of the critical quark mass with chemical potential in lattice
and continuum units as
\be
\label{c1prime}
\frac{a m^c(\mu)}{a m^c(0)} = 1 + \frac{c'_1}{a m^c(0)} (a \mu)^2 + ..., \qquad
\frac{m^c(\mu)}{m^c(0)} = 1 + c_1 \left( \frac{\mu}{\pi T} \right)^2 + ...
\label{c1}
\ee
then $c_1$ and $c'_1$ are related by
\be
c_1 = \frac{\pi^2}{N_t^2} \frac{c'_1}{a m^c(0)} + \left( \frac{1}{T_c(m,\mu)} \frac{d T_c(m^c(\mu),\mu)}{d(\mu/\pi T)^2} \right)_{\mu=0}.
\ee
Since $c_1'$ is consistent with zero, the second term dominates, leading
to an overall negative coefficient $c_1=-0.7(4)$ \cite{fp3}.
This is evidence that, in the $N_f=3$ theory on an $N_t=4$ lattice,
the region of first-order transitions {\em shrinks} as a baryon chemical potential
is turned on, and the unexpected scenario of \fig\ref{2schem} (right) is realized.
Interestingly, similar qualitative conclusions are obtained from simulations of the three flavour theory
with an isospin chemical potential~\cite{DKS_mc}, as well as simulations at imaginary $\mu$
employing Wilson fermions \cite{luo1}. 

\begin{figure}[t]
\vspace*{-0.5cm}
\includegraphics*[width=0.5\textwidth]{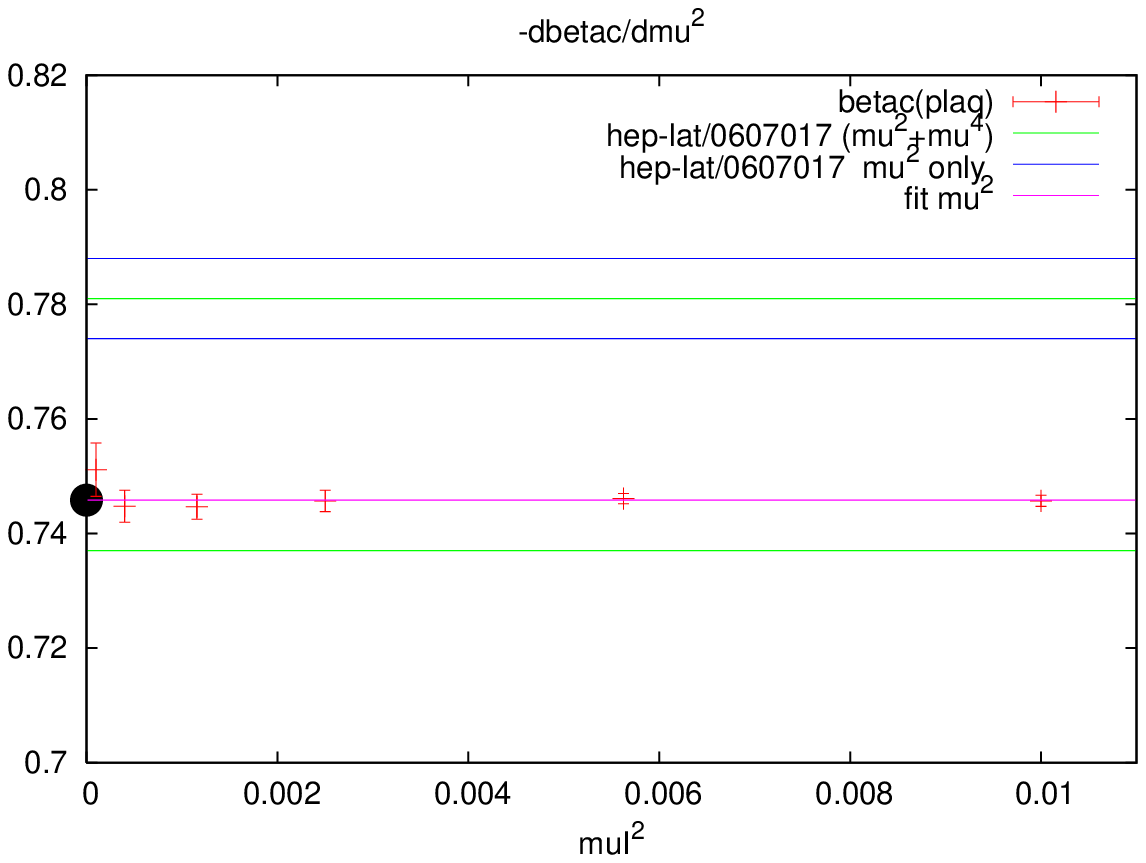}\hspace*{1cm}
\includegraphics*[width=0.5\textwidth]{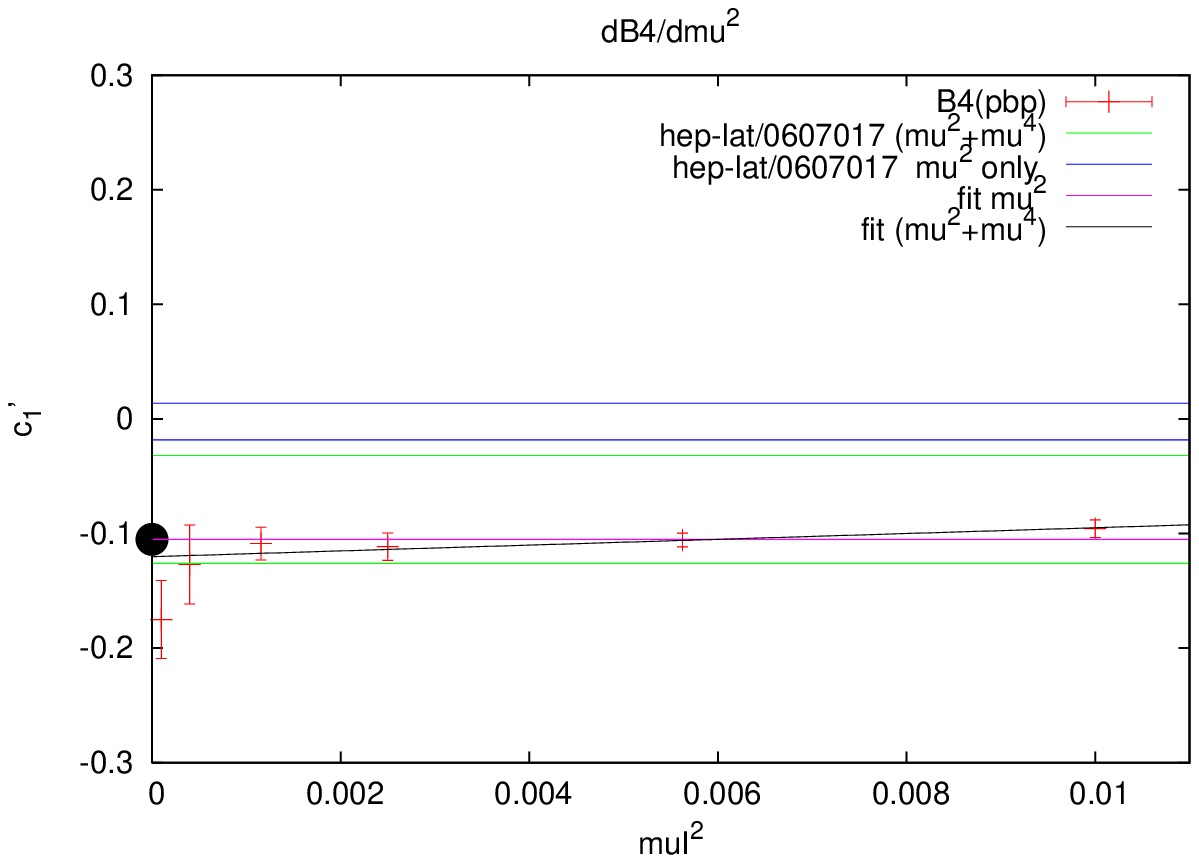}
\caption[]{Finite difference calculation of the $d\beta_c/d(a\mu)^2$ (left) and 
$c_1'\sim (\partial B_4/\partial (a\mu)^2$ (right). The error bands correspond to the leading and 
next-to-leading order fits from \cite{fp3}.}
\label{deriv}
\end{figure}

\subsection{Check of systematics}
\label{check}

Clearly, this is a rather consequential finding, and one must ask how reliable it is. Firstly,
our result for $c_1$ is only two standard deviations away from one with opposite sign. Second, one might
worry about systematic errors when fitting only the first term of the Taylor series.
Higher terms with alternating signs could fake a leading coefficient $c_1'\approx 0$ 
in a truncated series. Such terms would no 
longer cancel after continuation to real $\mu$, leading to a different picture.
In order to clarify this question we are currently performing a direct calculation of the derivatives
$d\beta_c/d(a\mu)^2$ and $\partial B_4/\partial (a\mu)^2$, thus avoiding a fit of the whole
function by a truncated Taylor series.
We compute the derivatives as finite differences, 
\be
\frac{dO}{d(a\mu)^2}=\lim_{(a\mu)^2\rightarrow 0}\frac{O(a\mu)-O(0)}{(a\mu)^2},
\ee
which offers several means to improve the signal. $O(a\mu)$ is evaluated by reweighting,
which has no overlap problem for $|a\mu|\leq 0.1$, thus eliminating uncorrelated 
fluctuations in the terms of the difference. The reweighting is done
in the imaginary $\mu$-direction, since then there is no sign problem and the reweighting factor
is less noisy. This in turn permits to compute the reweighting factor by noisy estimators rather 
than exactly, which is much cheaper. 

The results of these computations are shown in \fig\ref{deriv}. The derivatives are obtained by 
extrapolation to $\mu=0$: $d\beta_c/d(a\mu)^2=-0.746(1)$, 
$c'_1=-0.11(2)$. Also shown in the figures are the one sigma error bands from our previous  
$O(\mu^2)$ and next-to-leading $O(\mu^4)$ direct fits to $B_4$, as tabulated in \cite{fp3}. 
Note that the derivatives are completely consistent with the $\mu^4$ fits, but not with the 
leading order ones, i.e.~we are indeed sensitive to more than just the leading coefficient. 
In the derivative calculation this is 
indicated by the slight slope. Note, that the errors of the derivative computation are dramatically smaller,
so that now $c_1'$ is non-zero and negative beyond reasonable doubt. As discussed above,
the conversion to continuum units adds a dominant negative contribution to this.
We conclude that, on an $N_t=4$ lattice, the phase diagram \fig\ref{2schem} (right) is realized.

\section{Discussion}

Our results about the critical surface appear to be in qualitative contradiction 
with those of \cite{fk2}, \cite{ggpd}, which both conclude for the existence of a critical point
$(\mu_E,T_E)$ at small chemical potential $\mu_E/T_E \lsim 1$.  However,
in considering the reasons for such disagreement, one can see that the different data sets
are actually not inconsistent with each other, and the differing pictures can be explained
by standard systematic effects.

\begin{figure*}[t]
\vspace*{-1cm}
\centerline{
\scalebox{0.62}{\includegraphics{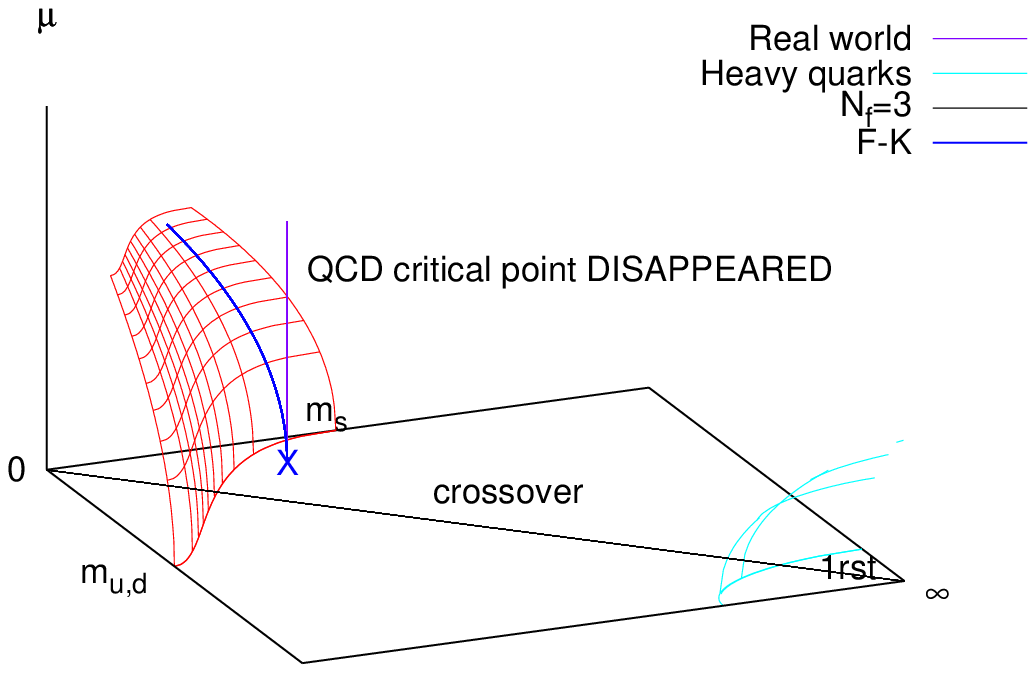}}
\scalebox{0.62}{\includegraphics{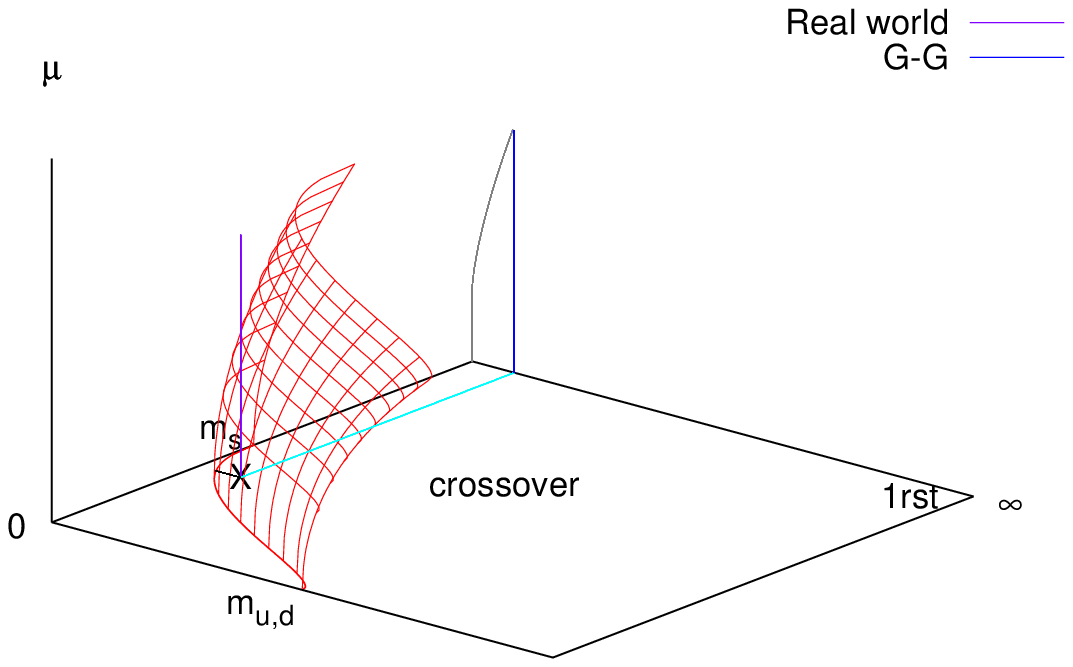}}
}
\caption{Left: Effect of keeping the quark mass fixed in lattice units in \cite{fk2}.
Right: Comparison at finite $\mu$ between the $N_f=2+1$ and the $N_f=2$ theory
considered in \cite{ggpd}.}
\label{syst}
\end{figure*}

In \cite{ggpd} the critical point was inferred from an estimate of
the radius of convergence of the Taylor expansion of the free energy.
Regardless of the systematics when only 4 Taylor coefficients
are available, the estimate is made for $N_f=2$. 
The $(\mu,T)$ phase diagram of this theory might well be qualitatively different
from that of $N_f=2+1$ QCD, as illustrated in \fig\ref{syst} (right). 
Its critical endpoint point, obtained 
by intersecting a critical surface when going up vertically from the 
$N_f=2$ quark mass values,
is clearly a long way from the critical endpoint of physical QCD, and nothing follows quantitatively
from the value of one for the other.

In \cite{fk2} the double reweighting approach was followed.
By construction, this reweighting is performed at a quark mass fixed in lattice 
units:
$a m_{u,d}=N_t\frac{m_{u,d}}{T_c} = const$. Since the critical temperature 
$T_c$ decreases 
with $\mu$, so does the quark mass. This decrease of the quark mass pushes
the transition towards first order, which might be the reason why a critical
point is found at small $\mu$. This effect is illustrated in the 
sketch \fig\ref{syst} (left), where the bent
trajectory intersects the critical surface, while the vertical line
of constant physics does not.
Put another way, \cite{fk2} measures the analogue of $c_1'$ instead of $c_1$
in \eq(\ref{c1}).
From their Fig.~1 \cite{fk2} we see that the distance of the theory from
criticality stays constant for small $\mu$, consistent with a small or zero 
coefficient $c'_1\approx 0$.
Taking the variation $a(T(\mu))$
into account could then make a dominant contribution, which might possibly change the results qualitatively.

A robust finding is the high quark mass sensitivity
of the critical point: irrespective of the sign, if 
$c_1\sim O(1)$ in \eq(\ref{c1}), then 
$m^c(\mu)$ is a slowly varying function of $\mu$, just as the pressure, screening
masses or $T_c$. Hence, $\mu_E(m)$ is rapidly
varying. A change of quark masses by a few percent will then imply
a change of $\mu_E$ by $O(100\%)$, which makes an accurate determination of $\mu_E$
a formidably difficult task. Our combined results of a negative curvature of the
critical surface as well as an increasing gap between the $\mu=0$ critical line
and the physical point make the existence of a critical point at small $\mu_E$
very unlikely.

A last note of caution concerns the fact that most investigations have used unimproved 
staggered quarks on coarse $N_t=4$ lattices only. This might be worrisome
given the exceedingly light quarks involved, as one should take the continuum limit
before the chiral limit for this discretization \cite{sharpe,gss}. Investigations on finer lattices
are indispensable in order to settle these issues.
Finally, a more complicated phase structure with additional
critical surfaces in \fig\ref{2schem} might allow for a critical point not continuously 
connected to the $\mu=0$ transition.\\

\noindent
{\bf Acknowledgement:} Unpublished work reported here is done in collaboration
with P.~de Forcrand.

\end{document}